# Excellence networks in science:

# A Web-based application based on Bayesian multilevel logistic regression (BMLR) for the identification of institutions collaborating successfully

Lutz Bornmann[1], Moritz Stefaner[2], Felix de Moya Anegón[3], & Rüdiger Mutz[4]

1       Division for Science and Innovation Studies, Administrative Headquarters of the Max Planck Society, Munich, Germany

2       Eickedorfer Damm 35, 28865 Lilienthal, Germany

3       CSIC, Institute of Public Goods and Policies (IPP), Madrid, Spain

4       Professorship for Social Psychology and Research on Higher Education, ETH Zurich, Zurich, Switzerland

**Abstract**

In this study we present an application which can be accessed via www.excellence-networks.net and which represents networks of scientific institutions worldwide. The application is based on papers (articles, reviews and conference papers) published between 2007 and 2011. It uses (network) data, on which the SCImago Institutions Ranking is based (Scopus data from Elsevier). Using this data, institutional networks have been estimated with statistical models (Bayesian multilevel logistic regression, BMLR) for a number of Scopus subject areas. Within single subject areas, we have investigated and visualized how successfully overall an institution (reference institution) has collaborated (compared to all the other institutions in a subject area), and with which other institutions (network institutions) a reference institution has collaborated particularly successfully. The "best paper rate" (statistically estimated) was used as an indicator for evaluating the collaboration success of an institution. This gives the proportion of highly cited papers from an institution, and is considered generally as an indicator for measuring impact in bibliometrics.

# 1  Introduction

In modern science, it has become commonplace for scientists to work together: "There is abundant evidence that research collaboration has become the norm in every field of scientific and technical research" (Bozeman, Fay, & Slade, 2013, p. 1). In most subject areas, individuals are no longer able to produce high-quality research without the help of their fellow researchers. According to Ziman (2000), Simonton (2013) and Cimenler, Reeves, and Skvoretz (2014) there are three main reasons for this: (1) Because many research projects need expensive equipment and data which is not available to everyone, willingness to collaborate is increasing. (2) It is only possible to solve many of the problems on which researchers work with an interdisciplinary approach, so researchers from different disciplines come together into teams. (3) For complex research subjects (such as climate research), it is essential to integrate researchers from different institutions and countries in one project. In the general view of Adams (2012), scientific knowledge is processed and combined more successfully in collaboration.

Today, bibliometrics is the most important method with which to evaluate research collaboration. The use of publications as a bibliometric measurement is based on the fact that researchers normally publish their interim and final results in scientific studies (Mulligan & Mabe, 2006; Smith, 1988). As the authors who are responsible for the content and their addresses are named on every publication, many studies have measured collaborations at author, institution and country level (see the overview in section 2). However, the co-authorship concept of measuring collaborations has also been criticised in recent years: Not all forms of collaborations between scientists lead to co-authorships (e.g. collaborations can simply consist of informal discussions between colleagues of the same institution) and scientists appear as co-authors on publications without substantial contributions (key word: honorary authorship). Katz and Martin (1997) discuss numerous possible cases where collaboration is not validly reflected in co-authorships. Despite the critique, co-authorship data has been the most frequently used data to measure collaboration in science. According to Bozeman et al. (2013) "the co-author concept of collaboration has several advantages, including verifiability, stability over time, data availability and ease of measurement" (p. 2).

This study presents an Web-based application (www.excellence-networks.net) with which it is possible to investigate national and international collaboration activities by institutions (see Chinchilla-Rodriguez, Vargas-Quesada, Hassan-Montero, Gonzalez-Molina, & Moya-Anegon, 2010). It compares institutions with each other in terms of their citation impact from co-authorships. This comparison consists of both a numerical (statistical parameters) and a graphical (network visualization) element. Accordingly, the application presents the results of graphical and statistical modelling of network data. It offers the following advantages compared to approaches of previous studies (see an overview in section 2): (1) We used an advanced statistical modelling approach for analysing the data. For example, this approach allows the calculation of credible intervals. (2) Whereas many previous studies presented their results as static co-authorship networks in publications, we developed a Web-based application and the user of the application (the reader of this publication) can inspect the most interesting results on collaboration. (3) Most of the Web-based applications visualizing institutional bibliometric data present the data for each institution separately (see e.g. the Leiden Ranking). Our application visualizes the institutions within an entire subject category and shows the performance of their collaboration activities.

The manuscript is organized as follows: After a literature overview of studies which have investigated collaboration in research (section 2), the used dataset for the Web-based application is described. It follows presentations (1) of the regression models which have been used to analyse the data (section 4) and (2) of the Web-based application which visualizes the results of the regression models (section 5).

## 2 Collaboration in research: literature overview

In recent years, numerous studies have been published which have used bibliometric data to look at collaboration in research. There is an overview of these studies in Katz and Martin (1997) and Bozeman et al. (2013). These studies show that, generally speaking, the probability of collaboration increases with closer physical proximity (see e.g. Katz, 1994). Even if there are activities, such as those in the European Union for example, with which to overcome national borders, the probability of collaboration does not increase (Chessa et al., 2013). Face-to-face discussions between scientists still seem even today to be an important factor in research (Ma, Fang, Pang, & Li, 2014). Furthermore, similarity of cultural and linguistic environments in collaborations plays a significant part. "Nigeria, for example, collaborates not with its neighbours in West Africa but with co-linguists in East Africa. This mirrors a global tendency to use paths of least resistance to partnership, rather than routes that might provide other strategic gains. Such language links have historically benefited the United Kingdom through alliances with Commonwealth countries that speak English and have adopted similar research structures" (Adams, 2012, p. 336). After all, "language, funding, intellectual property rights are country-dependent and constrain interaction between institutions" (Apolloni, Rouquier, & Jensen, 2013, p. 1468). The Web-based application introduced here can be used to observe specific patterns of institutional collaborations across national borders.

Even if similarity and physical proximity are important conditional factors in collaboration, the results of bibliometric studies indicate that there is more and more international collaboration between researchers which is being conducted over increasingly larger distances (Wagner, Park, & Leydesdorff, 2015). In one of the most comprehensive studies involving more than 21 million publications from almost every country and discipline (data from the Web of Science, WoS, from Thomson Reuters), Waltman, Tijssen, and van

Eck (2011) looked at the globalization of science by analysing co-authorships. As their results show, not only has the number of authors per publication risen steadily since 1980, but collaboration has become increasingly international. Similar results have been published by Larivière, Gingras, Sugimoto, and Tsou (2015). As results from Leydesdorff, Wagner, Park, and Adams (2013) show, furthermore, not only is there an observable increase in international collaborations over ever-growing distances, but also a change in the pattern of collaboration: while in the past collaborative work was dominated by (a few) European countries and the USA, today it involves a much larger group of around 50 countries.

Several bibliometric studies have already addressed the question of in how far collaboration represents a benefit to science. Most of these studies have examined whether collaboration has any effect on the impact of publications (measured in citations as one of the aspects of their quality). As the overviews of these studies by Sugimoto (2011) and Frenken and Hoekman (2014) show, we can expect publications produced in collaboration to have more citation impact than those which were not. The current study validates this finding. However, the results can vary depending on which country is being investigated and which citation indicator is used (Lancho-Barrantes, Guerrero-Bote, & Moya-Anegón, 2013; Levitt & Thelwall, 2010). It is countries with publications that achieve comparatively little impact (on average) which seem though to gain a lot from collaboration. The Web-based application introduced here allows inspecting country-specific patterns in more detail.

Other studies have also shown that working collaboratively can have a positive effect on the productivity of research units, on the granting of funding for research and overall on the generation of scientific knowledge (Subramanyam, 1983). "In the case of collaboration's effects on profits, wealth and economic development, the models tend to be more complex and over determined, but here too the preponderance of evidence is that research collaboration has salutary effects" (Bozeman et al., 2013, p. 4). Regarding a general cost-benefit evaluation of international collaboration, Adams (2012) writes: "So what are the costs and benefits of collaboration? It provides access to resources, including funding, facilities and ideas. It will be essential for grand challenges in physics, environment and health to have large, international teams supported by major facilities and rich data, which encourage the rapid spread of knowledge … As for costs, collaboration takes time and travel and means a shared agenda" (p. 336).

## 3     Methods – dataset used

The excellence-networks.net application is based on papers (articles, reviews and conference papers) published between 2007 and 2011. It uses the same data on which the SCImago Institutions Ranking (SIR, www.scimagoir.com) is based (data from Scopus, Elsevier). In this study institutional networks have been generated for a number of Scopus subject areas. For a subject area, co-authorship networks were generated for those institutions which published at least 500 papers in the publication period and which were also included in our tool excellence-mapping (www.excellencemapping.net). We refer to these institutions in the following as "reference institutions". Institutions with fewer than 500 papers in a category are not included in the application as "reference institutions". These reference institutions have been selected in the SIR database on the highest aggregation level. Thus, for example, the Max Planck Society is not included with the single Max Planck Institutes, but as the whole organization.

For every reference institution included in excellence-networks.net, the collaborating institutions have been identified. Collaborating institutions are those which have co-authored publications with the respective reference institution. We refer to the collaborating institutions as "network institutions" in the following.

In this study, the "best paper rate" is used as an indicator to measure citation impact (Bornmann, Mutz, & Daniel, 2013; Bornmann, Stefaner, de Moya Anegón, & Mutz, 2014a). This is the number of papers which are in the 10% of the most cited publications in their subject area and publication year.[1] The best paper rate is equivalent to the $PP_{(top\ 10\%)}$ in the Leiden Ranking (Waltman et al., 2012) and the excellence rate in the SCImago Institutions Ranking (Bornmann, de Moya Anegon, & Leydesdorff, 2012). The main variables for the statistical analysis are (1) the number of highly-cited papers produced by a reference institution in collaboration with a network institution and (2) the total number of papers they have published jointly. From the data (1) the best paper rates at the level of individual network institutions, which have cooperated with a certain reference institution and (2) the best paper rate for a reference institution, resulting from collaboration with all the network institutions are statistically estimated.

The citation window for the impact measurement in this study was from publication year to 2014. Only those network institutions producing at least 10 papers in collaboration with the reference institution have been taken into account in the statistical analyses. This restriction is necessary to ensure reliable statements about the extent and quality (impact) of collaborations.

To be able to show a reasonable minimum number of institutions in each subject area, only the results for those subject areas are shown in the application for which are at least 50 reference institutions. Table 1 lists the number of reference institutions and the mean number of network institutions for each reference institution for each subject area addressed in the application. The number of reference institutions ranges from 81 institutions for Psychology to 1279 for Medicine. The table shows the mean best paper rate besides the number of institutions. This rate is the average citation impact which the reference institutions have achieved in a subject area in collaboration with the network institutions in question.

---

[1] If one wishes to use the top 10% most frequently cited papers as an institutional performance indicator (Waltman et al., 2012), the tying of the ranks at the 10% threshold level can generate an uncertainty. SCImago introduces a secondary sort key in addition to citation counts as the solution for the problem of ranks tying at the threshold level: When the citation counts are equal, the publication in a journal with the higher SCImago Journal Rank (SJR2) (Guerrero-Bote & de Moya-Anegon, 2012) obtains the higher percentile rank. Adding this journal indicator takes into account not only the observed citations of the papers but also the prestige of the journals in which the papers are published. The results of Bornmann, Leydesdorff, and Wang (2013) demonstrate the advantage of the SCImago approach against other approaches of handling ties (e.g. fractional counting proposed by Waltman & Schreiber, 2013).

In total, the application is based on n = 460,144 network institutions related to the corresponding reference institutions.

Table 1. Number of reference institutions and average best paper rate for each subject area

| Subject area | Number of reference institutions | Average number of network institutions for each reference institution | Average best paper rate |
|---|---:|---:|---:|
| Agricultural and Biological Sciences | 570 | 35.0 | 0.25 |
| Biochemistry, Genetics and Molecular Biology | 800 | 69.0 | 0.35 |
| Chemical Engineering | 178 | 12.8 | 0.18 |
| Chemistry | 537 | 22.8 | 0.19 |
| Computer Science | 402 | 18.7 | 0.21 |
| Earth and Planetary Sciences | 361 | 105.1 | 0.33 |
| Engineering | 662 | 23.5 | 0.18 |
| Environmental Science | 267 | 34.7 | 0.30 |
| Immunology and Microbiology | 238 | 38.6 | 0.27 |
| Materials Science | 442 | 21.6 | 0.19 |
| Mathematics | 396 | 123.6 | 0.35 |
| Medicine | 1279 | 78.5 | 0.38 |
| Neuroscience | 132 | 40.0 | 0.34 |
| Pharmacology, Toxicology and Pharmaceutics | 102 | 18.2 | 0.24 |
| Physics and Astronomy | 698 | 166.9 | 0.45 |
| Psychology | 81 | 21.6 | 0.28 |
| Social Sciences | 229 | 11.4 | 0.27 |

# 4 Regression model

## 4.1 Statistical procedures

From a statistical point of view, the main objective of the Web-based application is to establish (1) a ranking of reference institutions in a scientific field and (2) a ranking of network institutions within single reference institutions regarding the success of collaboration. The application enables the user to check whether there are meaningful differences in collaboration activities between the institutions, i.e. beyond random.

*Basic statistical concept*

Bibliometric data is assumed to be exposed to random fluctuations (e.g., small changes due to the update of databases, coverage problems, and sampling errors), which require a statistical approach. Furthermore, the underlying network data for the Web-based application is hierarchically structured within the network. A single paper is nested within a network institution, which is itself related to a certain reference institution. For each paper the binary information is known, whether it belongs to the 10% most frequently cited papers or not. We used as underlying data for the application aggregated values for each network institution of a reference institution: the total number of co-authored papers and the number of co-authored papers belonging to the 10% most frequently cited papers (best paper rate). The best paper rate and its aggregate across all papers of a network institution, respectively, can be modelled by a logistic regression, where the frequencies (probabilities) are transformed to logits ($\log(p/(1+p))$) varying between minus to plus infinity as a basis for the application of the ordinary regression framework.

Bibliometric data from the same network institution related to a certain reference institution is more homogeneous than data between network institutions and reference institutions. Such measurement dependencies are considered in multilevel statistical models, which includes besides the ordinary regression model (mentioned above) the variability of data within different levels by variances (for level 1 the error variance is constant, $\pi^2/3$). If there is no systematic variability between network institutions and (especially) between reference institutions beyond random fluctuations, any comparison of institutions is useless (i.e., random samples of institutions).

*Bayesian approach*

For the statistical analysis in this study a Bayesian approach is favoured over a classical frequentist approach for the following two reasons: (1) Bayesian models can better deal with more complex data structures and huge data sets. (2) With Bayesian inference (e.g., credible intervals) some problems of classical null hypothesis significance testing (NHST) can be avoided. Bayesian statistical inference represents a learning process. Some prior knowledge either precise ("informative prior") or vague ("non-informative prior") about the modelled relationships is updated by the empirical estimation process ("likelihood"). This is similar to the usual concept in statistics ("frequentist statistics") not to obtain a single parameter estimation as usual but a parameter distribution, which represents the distribution of all possible parameter values given the prior information and the data. This posterior distribution should be more precise than the vague prior information (i.e., the prior distribution). For each parameter of the model and for each network and reference institution a posterior distribution can be estimated. In line with a more pragmatic interpretation of Bayesian statistics, credible intervals from the posterior distribution are used in this study to make a decision, whether a parameter value is of real importance or not. We used the term "statistical significant" in quotation marks as a simplified aid for the user of the Web-based application. Such a term makes in the realm of Bayesian statistics more sense than in the realm of ordinary frequentist statistics.

Following Bornmann, Mutz, Marx, Schier, and Daniel (2011), Bornmann, Mutz, et al. (2013), and Bornmann et al. (2014a), we prefer a multilevel approach for ego-centric individual network data in this study. According to Snijders, Spreen, and Zwaagstra (1995) the following conditions apply to this kind of data:

- the dependent variable (the best paper rate) is at the lowest level;
- the data contains no overlap of individual networks of different reference institutions or at least it was possible to ignore this overlap;
- the data yielded from different reference institutions (or egos) is mutually independent.

Given the network data at hand, it is clear that these are rather strong assumptions. A reference institution (ego) is connected to various network institutions. But a network institution can also be a reference institution. Therefore, the data obtained from different reference institutions is not fully independent. Multiple co-authorships with authors from different institutions create duplicates of papers in the data, in as much as different authors and different institutions contribute to a paper. Such duplicates usually occur if any kind of fractional counting of papers, i.e. transformations of raw data (Waltman & van Eck, 2015), is avoided. Measurement dependencies are taken into account to some extent by a multilevel statistical modelling strategy. The higher the measurement dependency, either for numerical (the same paper with different network institutions) or empirical reasons (empirical similarity of network institutions in their performance), (1) the more the number of independent units decreases, (2) the higher the differences between the respective mean value of an institution and the overall mean value are, and (3) the more the parameter of an institution shrinks statistically to the overall mean value across all institutions in a subject area. In statistics such estimates are also called "Empirical Bayes estimates" (Greenland, 2000; Hox, 2010).

*Bayesian multilevel logistic regression approach*

In this study a Bayesian multilevel logistic regression approach (BMLR) for binary outcomes was preferred, which takes into account the hierarchical structure of data and properly estimates the parameters and accuracy intervals (Bornmann, Mutz, et al., 2013; Congdon, 2010; Mutz & Daniel, 2007; Snijders & Bosker, 2004). Additionally, the model can handle complex network data structures with a small set of parameters. For example, one parameter is sufficient to test statistically whether the institutional best paper rate varies only randomly or in a systematic manner. Rankings among institutions only make sense if there are differences between these institutions that cannot be attributed to random fluctuation. In contrast to the method used by Bornmann, Stefaner, de Moya Anegón, and Mutz (2014b) for excellencemapping.net, a Bayesian model version was favoured over an ordinary frequentist approach for several reasons (Congdon, 2010; SAS Institute Inc., 2013):

The huge sample size of the data makes it impossible to estimate the parameters with ordinary multilevel models. Furthermore, a Bayesian approach is not only able to handle quite complex data structures, but it also allows whole parameter distributions (not fixed parameters) to be estimated for each reference institution and the related network institutions. Here, the mean of the parameter distribution represents the expected value of the posterior distribution which is of most interest for the visualization of the collaboration success of an institution in the Web-based application. The standard deviation and credible interval, respectively, gives some information about the uncertainty of the parameter estimation. Credible intervals make direct probability statements possible. For example, a 90% credible interval of 1.5 and 3 means that the true value of a parameter lies within this interval with the probability of .90 (Jebb & Woo, 2015). Credible intervals can be calculated for all model parameters.

Whereas in the frequentist approach a parameter is treated as a single fixed value or constant, in the Bayesian approach a parameter is treated as a random variable with a specific parameter distribution (Gelman et al., 2014). According to the Bayes theorem, the central idea of Bayesian statistics is to start with a prior distribution for the parameter in question $p(\theta)$, which represents the beliefs or uncertainty about the parameter prior to the empirical study (mean, skew, …). If there is no specific information about a parameter, a so-called non-informative prior with maximum uncertainty is chosen which does not strongly affect the estimation process unlike an informative prior. Non-informative prior is "intended for use in situations where scientific objectivity is at a premium for example, when presenting results to a regulator or in a scientific journal, and essentially means the Bayesian apparatus is being used a convenient way of dealing with complex multi-dimensional models" (Lunn, Jackson, Best, Thomas, & Spiegelhalter, 2013, p. 81). The posterior estimates are then quite similar to the estimates obtained by maximum likelihood. This way of thinking is in line with the concept of "statistical pragmatism" of Kass (2011). Note that many varieties of Bayesian statistical analysis exist (e.g., a strong definition of subjective prior probabilities versus neglecting the problem of prior definitions), which nevertheless strongly agree on the basic ideas of Bayesian inference (see https://bayesian.org/Bayes-Explained). In the light of the empirical data, which is formalized by the likelihood of the statistical model, p(y|θ), and the prior, p(θ), the posterior distribution of the parameter p(θ|y) is updated and estimated: p(θ|y) ∝ p(y|θ) p(θ).

*Bayesian multilevel logistic regression*

In this study, a BMLR model is assumed (Hamaker & Klugkist, 2011), which is defined on three levels. Papers are clustered within network institutions and network institutions are clustered within reference institutions, where $j$ ($j = 1 \ldots$ R) denotes the reference institutions or level-3 units, $i$ ($i = 1 \ldots$ N) denotes the network institutions or level-2 units, and $r$ ($r = 1 \ldots K_{ji}$) indicates the level-1 units ("papers"). The expectation is that papers from the same reference or network institutions are likely to be cited more homogeneously than papers from different institutions. The dependent variable $y_{jir}$ is binary (1 = paper $r$ belongs to the 10% most frequently cited publications, 0 = paper $r$ does not belong to the 10% most frequently cited papers). In order to simplify the analysis we use aggregated values: the total number of highly cited papers for each reference institution in collaboration with a network institution, $y_{ji}$, which is a subset of all the papers produced in collaboration, $n_{ji}$, and is binomially distributed ("~" for distributed in Eq. 1). The estimated probability, $p_{ji}$, is the proportion of highly cited papers a reference institution $j$ has published with another network institution $i$. We call this proportion the "best paper rate". Thus, the best paper rate in this study always relates to the impact which at least two institutions have achieved in collaboration.

The BMLR model for each subject area can be formalized as a three-level model (Bornmann et al., 2011; Gelman & Hill, 2007; Goldstein, 2011; SAS Institute Inc., 2011; Snijders et al., 1995):

Level 1:

$$y_{ji} \sim \text{binomial}(p_{ji}, n_{ji}) \tag{1}$$

Level 2:

$$p_{ji} \sim \text{logistic}(u_{0ji}) \qquad u_{0ji} \sim N(\tau_j, \sigma^2_u)$$

Level 3:

$$\tau_j \sim N(\beta_0, \sigma^2_\tau),$$

where $\sigma^2_u$ denotes the mean variance of the random effects $u_{0ji}$ between the institutions in the network for a reference institution (level 2, *within variance*), and $\sigma^2_\tau$ denotes the variance of the random effects $\tau_j$ between the reference institutions themselves (level 3, *between variance*). The term $\beta_0$ denotes the overall intercept or average impact in a subject area. By hierarchical centring (Lunn et al., 2013) the expected values of the random effects of level 2 are the random effects of level 3 units, and precisely not zero, as it is the case in an ordinary frequentist multilevel model. This procedure enhances the convergence of the Monte Carlo simulation for estimating the parameter distributions.

The following non-informative priors are assumed, where N denotes the normal distribution and N+ the half-normal distribution:

$$\beta_0 \sim N(0, \text{var} = 1{,}000) \tag{2}$$
$$\sigma^2_u, \sigma^2_\tau \sim N^+(0, \text{var} = 1{,}000)$$

There is a so-called intra-class correlation, $\rho = (\sigma^2_u + \sigma^2_\tau)/(3.29 + \sigma^2_u + \sigma^2_\tau)$, which reflects the homogeneity of papers within a reference institution, where $\sigma^2_u$ and $\sigma^2_\tau$ are the respective means of the corresponding parameters of the posterior distribution. The number 3.29 represents the constant error variance ($=\pi^2/3$). For example, an intra-class correlation of zero means that the reference institutions and the network institutions differ only by random. In that case there is no homogeneity of citation data within network institutions or reference institutions.

In order to check for real differences or effects (e.g., parameter differs from zero), a rather simple approach is to use the posterior distribution of a parameter itself. Eventually, this is the empirical update (likelihood) of the priors, and represents the probability density of all parameters that are possible given the data and the vague priors or p(H0|D), NOT p(D|H0) "to state the problem in Bayesian terms (where *p*-values are about the plausibility of the hypothesis given the data, as opposed to the other way around)" (Gelman & Loken, 2014, p. 461). As an analogue to statistical significance testing in frequentist statistics it can be checked whether the credible interval includes zero or not. If not, there is a "statistically significant" effect, i.e. an effect beyond random fluctuation. Kass (2011) speaks of a "pragmatic interpretation of posterior interval" (p. 4) in this context. Jebb and Woo (2015) coin the term "practical significance" (p. 11).

In this study, the term "statistical significance" is adopted from frequentist statistics. At the first glance, its use in Bayesian statistics might be problematic. However, we stick to this term for the following three reasons: (1) It is common to use the same statistical terms in Bayesian and frequentist statistics but with very different meanings. A good example is the term "probability": Whereas probability is defined as a long run frequency in frequentist statistics, it is defined as a measure of degree of subjective beliefs about the values of a parameter in Bayesian statistics. (2) Even in the statistical literature on Bayesian statistics frequentist concepts, as e.g. p-values, are discussed. Bayarri and Berger (2004) speak of a "posterior predictive p-value" following classical NHST reasoning. Thus, the concept of statistical significance is controversial, but not wrong in Bayesian statistics. (3) Our Web-based application requires easy and intuitive understandable concepts for users not familiar with statistics.

With the BMLR model shrinkage it is possible to obtain estimates similar to the Empirical Bayes estimates (EB) in ordinary multilevel modelling which are more precise than their empirical counterparts, the raw probabilities (Bornmann, Mutz, et al., 2013; Hox, 2010; SAS Institute Inc., 2011).

*Visualization of the regression results*

For the visualization of the networks the following indicators were calculated. The logistic transformed random effect, logistic ($\tau_j$), especially the means of the corresponding parameter distributions, provides for the *best paper rate of the reference institution*. To obtain a value for the mean best paper rate of a network institution, first the predicted number of highly cited papers in a subject area, $y_{ji}$, and its parameter distribution were estimated. Afterwards, the mean of this parameter distribution is divided by the total number of papers produced in collaboration between a reference institution and the respective network institution, $n_{ji}$, in order to obtain the *best paper rate of the network institution*. Predicted values "contain samples from the posterior predictive distribution of the response variable" (SAS Institute Inc., 2011, p. 4314). The overall best paper rate of a subject area is also derived from these predicted values. The multiplication of the standard deviation of the parameter distribution for each reference institution or network institution by 1.39 instead of 1.96 results in so-called Goldstein-adjusted credible intervals (Goldstein & Healy, 1995) with the property that if the credible intervals of two institutions do not overlap, they would differ "statistically significantly" ($\alpha = 5\%$) in their estimates (i.e. best paper probabilities). We use inverted commas with "statistically significantly" because this term from classical frequentist statistics has been adopted for Bayesian analysis.

To generate the data for the visualization for each of the 17 subject areas separately (see section 3) a BMLR analysis was calculated using the highly cited papers (those written collaboratively). Due to the huge data sets for each area, and the complexity of the model, the data was randomly split into different subsets (if necessary). Subsequently, the statistical model was estimated for the different subsets to generate the interesting mean values of the parameter distribution, and the credible intervals.

*Software*

The analyses have been done using the PROC MCMC procedure implemented in the statistical software SAS 9.4 (SAS Institute Inc., 2011). A Metropolis-Hastings algorithm was used with 10,000 iterations and 1,000 burn-in samples. The MCMC simulation procedure kept every second simulation sample and discarded the rest (thinning = 2). The iteration process was time consuming; it amounted to about one day.

**4.2   Results of the Bayesian multilevel regression**

We chose the subject area "Pharmacology, Toxicology and Pharmaceutics" to illustrate the statistical approach which is identical for all subject areas (however, the example does not require any data splitting). It would go far beyond this paper to report the results of all 17 subject-specific analyses.

A model was taken as the starting point in the statistical analysis, which only includes the intercept $\beta_0$. The Deviance Information Criterion (DIC), which can be used to compare different Bayesian models for the same data, amounted to 10,278.98. This was markedly higher than the DIC for the full model (DIC = 8,190, Table 2), as formalized in Eq. (1). Due to the fact that the smaller the DIC, the better the model is, the full model was favoured over the null model. In other words the differences between network institutions within reference institutions, and the differences between reference institutions themselves were far from random.

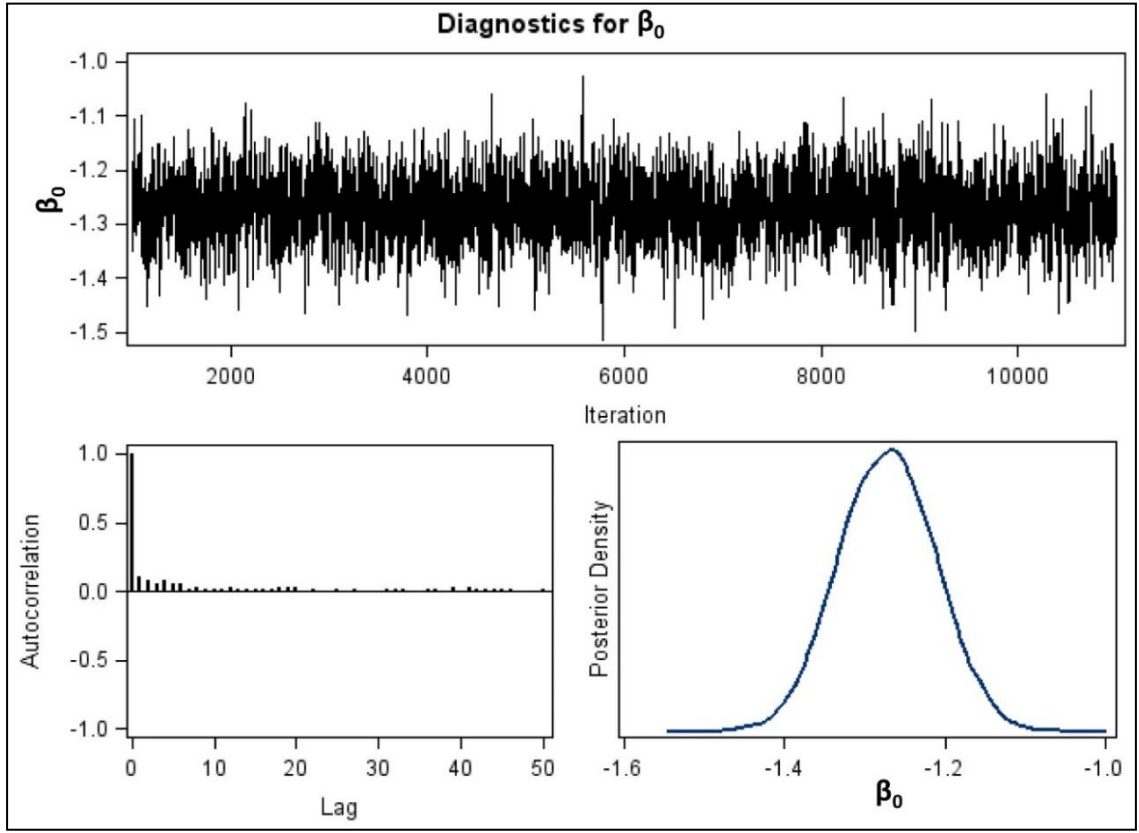

Figure 1. Diagnostic plots for the intercept $\beta_0$.

Before the model was finally estimated, we checked whether the chain in the Monte Carlo simulation converges. Convergence of the iteration process is of great importance to justify the later interpretation of the estimation results. In Figure 1, the posterior distribution for all parameters is shown with the mean, the standard deviation, and the highest posterior density 95% credible interval (HPD). Autocorrelation plots, trace plots, and posterior density plots for the intercept and the variance components, as shown in Figure 1 (and Figure 2), serve as diagnostics for the chain. Convergence is reached, if (1) the autocorrelation between successive iterations vanishes completely, (2) the trace plots show no systematic trends (random noise, stability of the chain), and (3) the posterior density of the parameter converges to a normal distribution. For both selected model parameters, the mixing of the chain looks quite reasonable and stable, suggesting convergence of the chain. The same is also true for the variance component $\sigma^2_u$ (Eq. 2).

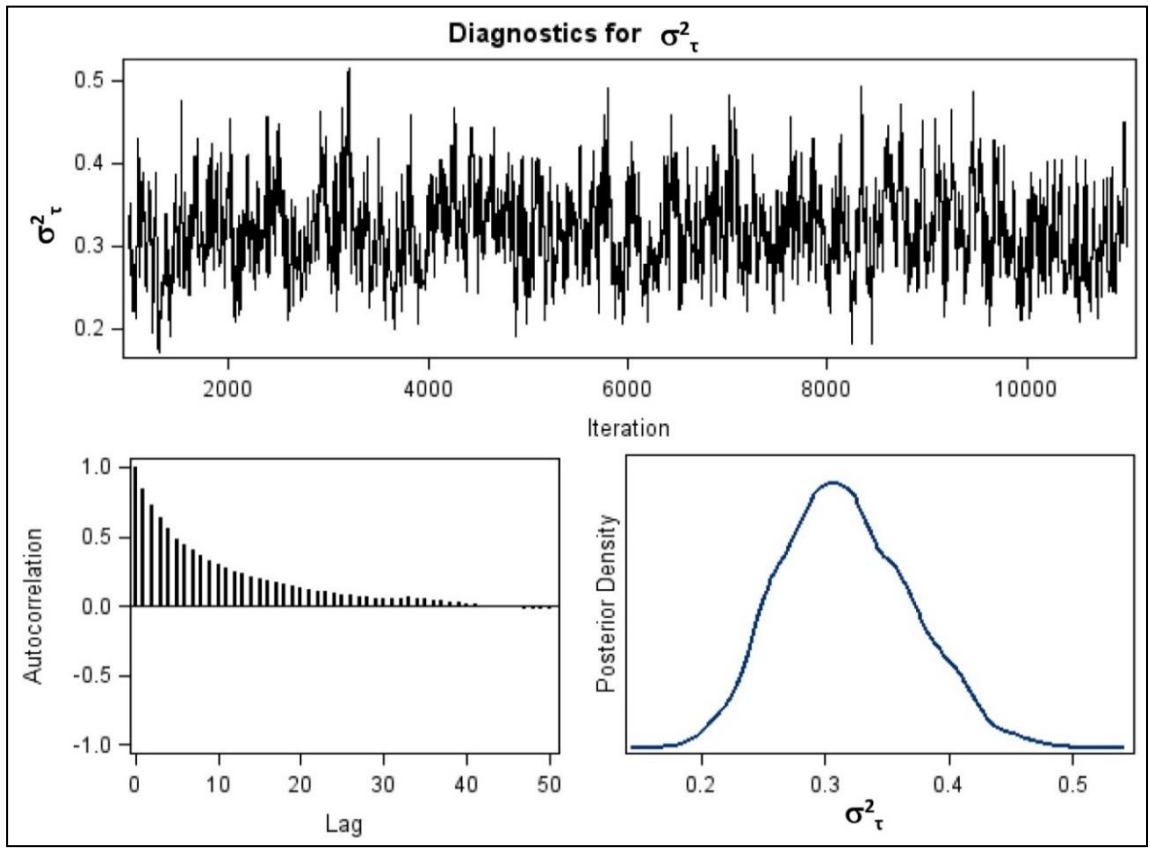

Figure 2. Diagnostic plots for the variance component $\sigma^2_\tau$

In order to test the parameters for "statistical significance" the HPD was chosen, which is defined as follows (SAS Institute Inc., 2013):

"A 100 (1-$\alpha$) % HPD interval is a region that satisfies the following two conditions:

1. The posterior probability of that region is 100(1-$\alpha$)%.

2. The minimum density of any point within that region is equal to or larger than the density of any point outside that region" (p. 133).

For all model parameters the HPD did not contain zero, which speaks for the "statistical significance" of the parameters. In other words, not only the network institutions differed within the reference institutions with respect to the best paper rate far from random fluctuations, but also the reference institutions differed systematically from one another. The variance component of the reference institutions (mean of the posterior distribution: $\sigma^2_\tau$ =0.32) is more than twice as high as the variance component of the reference institutions (mean of the posterior distribution: $\sigma^2_u = 0.14$). This demonstrates larger differences in the mean best paper rate on the level of reference institutions than on the level of nested network institutions. Eventually, rankings of institutions using the estimated best paper rate are possible both on the level of reference institutions and on the level of network institutions. According to the variance components, parameter distributions of each reference and network institution can be calculated. These parameters and the corresponding credible intervals are presented in our application.

The intra-class correlation [0, 1] as a measure of the variability between institutions to the total variability as the sum of the variability between and within institutions (random fluctuations) amounts to 0.12 [0.10; 0.15]. Overall, the correlation is rather low: The differences between the institutions in "Pharmacology, Toxicology and Pharmaceutics" are not as large. The mean of the parameter distribution of the intercept parameter $\beta = -1.27$ represents the overall probability or best paper rate in the subject category on a logistic scale. Transformed back to a probability ($p=\exp(-1.27)/(1+\exp(-1.27))$) the overall "best paper rate" amounts to 0.22. This value slightly differs from the value presented in the application ($p = 0.237$ or 23.7%, see Figure 4). The value in the application is calculated on the base of the predicted number of papers belonging to the 10% most frequently cited papers in a subject category and the overall number of papers in the subject category. The observed value for this category is 0.24 (see Table 1).

With respect to the other subject categories which we have considered in the application besides "Pharmacology, Toxicology and Pharmaceutics", the variance components are also large enough to justify the current specification of the model. The convergence of all models is also warranted. For "Psychology" the variability between reference institutions and network institutions, respectively, is rather low ($\sigma^2_\tau = 0.07$ $\sigma^2_u = 0.09$), but sufficient.

Table 2. Model estimation for "Pharmacology, Toxicology and Pharmaceutics" ($N_{\text{Net. inst.}}$ = 1,861, $N_{\text{Ref. inst.}}$ = 102)

| Variables | Parameter | Mean | SD | HPD Lower | HPD Upper |
|---|---|---|---|---|---|
| *Fixed effect* | | | | | |
| Intercept | $\beta_0$ | -1.27 | 0.06 | -1.39 | -1.15 |
| *Random effects* | | | | | |
| Level 2 "Net. inst." | | | | | |
| $u_{ji}$ | $\sigma^2_u$ | 0.14 | 0.01 | 0.11 | 0.17 |
| Level 3 "Ref. inst." | | | | | |
| $\tau_j$ | $\sigma^2_\tau$ | 0.32 | 0.05 | 0.22 | 0.42 |
| ICC | $\rho$ | 0.12 | 0.01 | 0.10 | 0.15 |
| DIC | | | 8,190 | | |

Note. HPD = highest posterior density 95% credible intervals, ICC = intra-class correlation, p = predicted best paper rate, Net. inst. = network institution, Ref. inst. = reference institution, DIC = Deviance Information Criterion

# 5 Web-based application

## 5.1 Visualization techniques and statistics

The network layout mode of the application indicates the connectivity structure of each subject area through spatial arrangement. This is achieved through running an optimization algorithm called "ForceAtlas 2" (Jacomy, Venturini, Heymann, & Bastian, 2014), as part of gephi 0.8.2 (http://gephi.org, which is open source software for network analysis). By running a physical simulation, more strongly connected nodes are pulled together more closely on the map. The algorithm is initialized with geographic coordinates. Being deterministic in nature, this initialization creates stable, reproducible results in the simulation. Further, it results in a layout where geographic topology is partly preserved (if compatible with network connectivity), thus allowing a more intuitive reading of the map. In the last step in network calculation, overlaps between nodes are removed, in order to facilitate readability. A screen recording of an example of an optimization procedure can be seen at https://vimeo.com/131653752.

Classical network analysis was applied on the data (Börner, Sanyal, & Vespignani, 2007; Knoke & Yang, 2008; Kumar, 2015). A 0/1-adjacency matrix was generated with 1 indicating collaborations between a reference and a network institution, and 0 indicating no collaboration. The R-package SNA of the R-software was used to calculate various network measures. To avoid information overload, we restricted reporting to only one measure, the "betweenness centrality". "The betweenness concept of centrality concerns how other actors control or mediate the relations between dyads that are not directly connected. *Actor betweenness centrality* measures the extent to which other actors lie on the geodesic path (shortest distance) between pairs of actors in the network" (Knoke & Yang, 2008, p. 67). The measure yields high values for institutions which have an important structural role in the network as centres of a large cluster or in connecting two separate clusters (for example). As our study investigates collaboration effects, betweenness centrality was chosen over simpler approaches, such as counting merely the number of connections between institutions (degree centrality).

In the network layout view of the application, the circle size of each institution corresponds to its betweenness centrality. However, when an institution is selected by the user of the application, the circle size represents the total amount of collaborations with it. Alternatively to the network layout view, users can choose to view institutions on a world map, using the van der Grinten projection (Snyder, 1926).

In both views, the colouring can be switched between two modes: "Colour by Country" colours the institutions based on the country they are located in. Colours are only chosen for category discrimination; any colour similarities are unavoidable, but coincidental. "Colour by best paper rate" colours the institutions based on the best paper rate. The diverging colour scale ranges from red (lowest value in the subject area) through grey (average value in the subject area) to blue (highest value in the subject area). Colours are interpolated in CIELCH colour space (Hunter, 1948), in order to achieve perceptual homogeneity along the value scale.

When an institution is selected by the user, the colour represents the best paper rate of the collaborations between the institution selected and other institutions. When extreme values might land outside the domain of the colour scale, clamping is applied, and the respective value is plotted with a colour corresponding to the subject area's minimum or maximum value.

The Web-based application is based on HTML5 techniques and was implemented using coffeescript, sass, gulp, bower, d3.js, backbone, underscore.js, and bourbon.

## 5.2 Explanation of the application

In the following, we present the Web-based application which is based on the results of the BMLR models described in section 4. For the presented institutional co-authorship networks the predicted values given the mean of the estimated posterior distribution of the corresponding parameters were used.

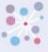

Figure 3. Start screen for www.excellence-networks.net

Figure 3 shows the start screen for the Web-based application. This page has an introductory explanation of the application and offers the option of choosing the excellence network for a certain subject area ("Select a subject area to start") or displaying a text which provides a more detailed overview of the data, the methodology and the visualization ("More information"). If the user selects a subject area, a new screen opens which displays all the reference institutions that have been taken into account for that subject area. We have selected as an example for the following explanations "Pharmacology, Toxicology and Pharmaceutics" – similar to the approach taken in section 4. The explanations can be applied to all the other subject areas.

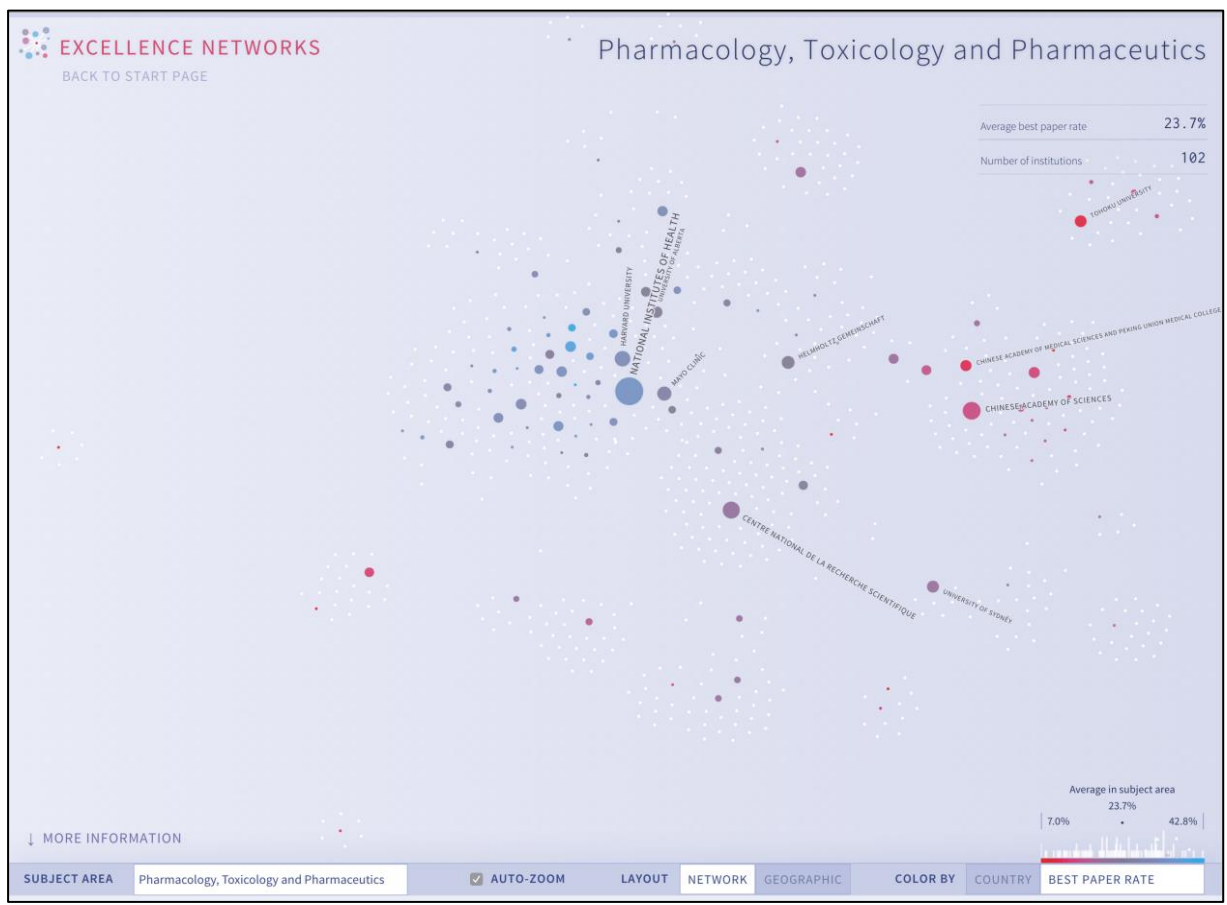

Figure 4. All reference institutions in the "Pharmacology, Toxicology and Pharmaceutics" area for which co-authorship networks can be displayed

Figure 4 shows the screen after "Pharmacology, Toxicology and Pharmaceutics" has been selected. Users will see the references institutions in this subject area which have published at least 500 papers from 2007 to 2011 in this field. The size of the circles is proportional to the betweenness centrality of an institution. Moving the mouse over an institution opens a window in which the name of the institution and the country in which it is located are displayed. It also gives the average best paper rate which the reference institution has achieved with its network institutions. The average best paper rate is predicted from the statistical model, especially from the average of the corresponding posterior parameter distribution. Users can choose between two different methods for arranging the institutions (see "Layout" at the bottom edge of the screen): (1) "Network" depends on the amount of collaboration between the institutions: The more frequently a reference institution has published highly-cited papers jointly with other network institutions, the more central its position. As Figure 4 shows, the National Institutes of Health occupies a particularly central position in "Pharmacology, Toxicology and Pharmaceutics". (2) The positions of the reference institutions are shown on a map of the world ("Geographic"). White dots are shown on both arrangements on the screen next to the reference institutions shown in colour. These dots are institutions which become relevant (and are then displayed in colour) if the network institutions collaborating with a reference institution are shown.

| Collaborations in PHARMACOLOGY, TOXICOLOGY AND PHARMACEUTICS | | | | Find an institution |
|---|---|---|---|---|
| INSTITUTION | COUNTRY | # OF PUBLICATIONS | | BEST PAPER RATE |
| Centre National de la Recherche Scientifique | FRA | ||| | 20.6% | |
| Institut National de la Sante et de la Recherche Medicale | FRA | || | 22.5% | |
| National Institutes of Health | USA | || | 33.5% | |
| Veterans Affairs Medical Centers | USA | || | 29.5% | |
| Harvard University | USA | | | 30.7% | |
| Chinese Academy of Sciences | CHN | | | 14.6% | |
| Partners HealthCare System | USA | | | 33.6% | |
| Assistance Publique Hopitaux de Paris | FRA | | | 19.8% | |
| Ministry of Education of the People's Republic of China | CHN | | | 13.7% | |
| Seoul National University | KOR | | | 19.2% | |
| Consejo Superior de Investigaciones Cientificas | ESP | | | 16.8% | |
| University of Toronto | CAN | | | 29.3% | |
| Universidade de Sao Paulo | BRA | | | 13.1% | |
| National Taiwan University | TWN | | | 17.5% | |
| University of North Carolina, Chapel Hill | USA | | | 31.3% | |
| Consiglio Nazionale delle Ricerche | ITA | | | 20.6% | |
| Utrecht University | NLD | | | 25.4% | |
| University of Pittsburgh | USA | | | 32.2% | |
| Universita degli Studi di Roma La Sapienza | ITA | | | 19.6% | |
| Shanghai Jiao Tong University | CHN | | | 11.9% | |
| Peking University | CHN | | | 15.5% | |
| University of California, San Francisco | USA | | | 38.0% | |
| University of California, San Diego | USA | | | 38.6% | |
| Columbia University | USA | | | 40.6% | |

Figure 5. List of the reference institutions in "Pharmacology, Toxicology and Pharmaceutics" (sorted by the number of publications)

The colour of the circles can have two different meanings, which the user can designate in the bottom right-hand corner of the screen: (1) If the user selects the "Country" setting, the circles are coloured according to the association with a country. (2) The other option is shown in Figure 4. If the user chooses "Best paper rate" the institutions which have collaborated very successfully with network institutions are shown in blue. Red indicates less successful collaboration in the period specified. Grey represents collaboration which is average for the subject area.

The average value for the subject area is shown on the screen with the "Average best paper rate". For "Pharmacology, Toxicology and Pharmaceutics" this is 23.7% for a total of 102 reference institutions (see Figure 4). The area above the "Best paper rate" button is used to visualize the distribution of reference institutions over the colour scale (showing the minimum, the average best paper rate, and the maximum). For example, this allows users to see for the different subject areas whether the distribution of the reference institutions over the impact scale is even or skewed.

The list of visualized institutions can be found below the graphical representation (by clicking on "More information" or scrolling). The "Pharmacology, Toxicology and Pharmaceutics" list shows the 102 reference institutions which have published at least 500 papers in this subject area (see Figure 5). The country and the average best paper rate achieved with all the network institutions are given for each institution. The reference institutions can be sorted by country and best paper rate. For each reference institution, the best paper rate is given with a credible interval. This means that firstly, the reliability of the estimated impact scores can be evaluated (the bigger the credibility interval, the less reliable the estimate) and secondly, an assessment can be made of whether the performance of two reference institutions (collaborating with their network institutions) is "statistically significantly" different: there is a statistically meaningful difference if the credible intervals do not overlap.

Users can click on an institution to select it from the list of reference institutions or from the graphical representation. This displays the collaboration network of the reference institution selected and (below) the list of the network institutions collaborating with it.

Figure 6 shows the collaboration network for University College London in the "Pharmacology, Toxicology and Pharmaceutics" subject area: all the network institutions with which this reference institution has published at least 10 papers jointly (n=21 institutions) are shown. University College London has achieved an average best paper rate of 30.8% over all the network institutions. The "Country" and "Best paper rate" buttons can be used to colour the network institutions to reflect the country to which they belong or the citation impact achieved in collaboration. The "Network" and "Geographic" buttons can be used to select the various ways of displaying the network (see above).

Three different best paper rates are given in a window for each network institution with information about the success of collaboration between University College London and a network institution. This window is shown when the user hovers the mouse over the institutional node (the same window appears if the user hovers the mouse over an institution in the table). The average best paper rate which the institutions have achieved as reference institutions with their network institutions is given for both institutions. The best paper rate for those papers which the two relevant institutions have published jointly is also shown. If a network institution is not a reference institution, there is no corresponding average best paper rate in this window.

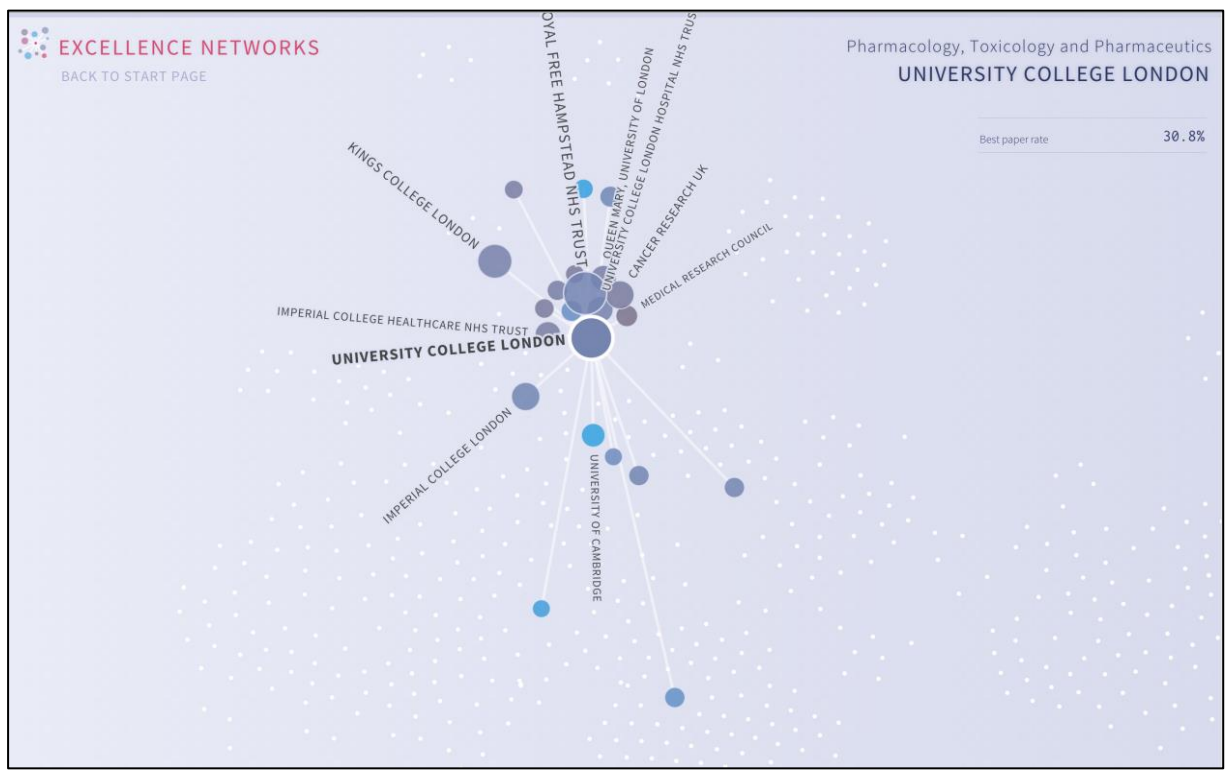

Figure 6. The collaboration network for University College London in the "Pharmacology, Toxicology and Pharmaceutics" subject area.

For example, as a reference institution, the University of Copenhagen has an average best paper rate of 31.5%; for University College London this rate is 30.8%. When the two institutions collaborate with each other, they have a best paper rate of 31.5%. Because at 30.8% the average best paper rate for University College London is slightly below the best paper rate achieved with the University of Copenhagen, University College London does not appear to benefit from the collaboration, if its success is measured using citation impact. As the average best paper rate for the University of Copenhagen is exactly the same as the best paper rate achieved with University College London, it too sees limited impact increase from collaborating. However, looking at the results for the National Institutes of Health, it can be seen that University College London benefits from this collaboration: with a best paper rate of 39.3% the papers produced collaboratively are cited much more frequently than the average best paper rate for University College London of 30.8% indicates.

In the network of a specific institution, users can click on the institutions that collaborate with it. However, this only works for those institutions which belong to the reference institutions in a subject area.

Below the graphical depiction of the network is a list of network institutions which cooperate with the reference institution selected. The list of institutions can be sorted by country, number of papers produced in collaboration and best paper rate (in collaboration). The credible intervals are also presented with the best paper rate. This allows "statistically significant" differences between the network institutions to be determined in terms of their collaboration with the reference institution. There is only a "statistically significant" difference between two network institutions when the credible intervals do not intersect. Otherwise the performance of the collaborations can be assessed as very similar.

# 6  Discussion

In this study we have presented an application which can be accessed via excellence-networks.net and which represents networks of scientific institutions worldwide. Applications of this kind which present a comprehensible visualization of complex data have grown in popularity over recent years: "Network and science-mapping visualizations have considerably enhanced the capacity to convey complex information to users. These tools are now sufficiently mature to be used not only in academia but also in consultancy and funding organizations" (Martin, Nightingale, & Rafols, 2014, p. 4). Frenken and Hoekman (2014) published an overview of the studies which have been concerned with these applications and the methods that underlie them. We are following this trend with our study and have enhanced another of our applications with which we visualize the performance of scientific institutions on a map and in ranked lists (excellencemapping.net) with a network application.

The networks which we have presented in this study are based fundamentally on collaboration measured by means of co-authorships. To examine collaboration by means of co-authorships, networks are generated on the basis of addresses noted on the publications (Bornmann & Leydesdorff, 2015). The generation of these networks is very popular nowadays. For example, Apolloni et al. (2013) have used network evaluations for the European Union to evaluate "under which geographical extent co-authorships have higher probability of resulting in high impact articles" (p. 1467). In this study, we have used co-authorship networks to examine how successfully overall an institution (reference institution) has collaborated (compared to all the other institutions in a subject area), and with which other institution (network institution) an institution (reference institution) has collaborated best. The "best paper rate" was used as an indicator for evaluating the success of an institution. This gives the proportion of highly cited papers from an institution and is considered generally as a robust indicator for measuring impact (Hicks, Wouters, Waltman, de Rijcke, & Rafols, 2015; Waltman et al., 2012).

The application presented here is based on a statistical model which amongst other things allows credible intervals of possible parameter values (in the light of the data – likelihood – and the prior information) to be calculated for the best paper rates in collaboration. With the aid of the credible intervals, users can assess whether the difference in performance of two institutions which collaborate with other institutions is "statistically significant". This is what distinguishes the application from the depiction of bibliometric results in university rankings in which the slightest differences in performance between institutions leads to differences in the ranking which are then interpreted as significant differences by users of the ranking (Hicks et al., 2015).

Our presentation of the results follows such initiatives as those of Waltman et al. (2012) which show the results of the Leiden Ranking with stability intervals. "A stability interval indicates a range of values of an indicator that are likely to be observed when the underlying set of publications changes" (http://www.leidenranking.com/methodology/indicators#stability-intervals). However, all statistical analyses, even the calculation of descriptive statistics for university rankings, require assumptions. Most critical is that the data used contains no overlap of individual networks for different reference institutions and the data yielded from different reference institutions (or egos) is mutually independent. In line with Snijders et al. (1995) multilevel modelling is the method of choice to cope with this problem, because the dependency in the data is explicitly taken into account, especially in the estimation of parameters (Hox, 2010).

A number of releases of the Web-based application excellencemapping.net have been issued in recent years (Bornmann et al., 2014a, 2014b; Bornmann, Stefaner, de Moya Anegón, & Mutz, 2015), which is similar to excellence-networks.net. The releases not only read in up-to-date data, but have also improved the methodology. For example, a covariance-adjusted ranking has been implemented in the application which means that certain contextual factors in research (such as the corruption perceived in a country) can be kept constant for all the institutions and the result of the performance estimates can be visualized under these consistent conditions. We also plan to optimize the first version of excellence-networks.net presented here and to make methodological improvements. We are therefore very interested in feedback from the users.

Generally speaking, users of the excellence-networks.net should only compare performance values from this application with each other and not with those published in other applications (such as the Leiden Ranking). The best paper rates shown in our application always relate to papers produced in collaboration (and this is normally not the case in other applications). As we showed in the introduction, better performance can be assumed for these papers than for papers which have not been produced in collaboration. When interpreting best paper rates it is normally possible to work with an expected value: 10% of publications from an institution can be expected to be among the 10% most cited publications. As it is possible to assume fundamentally a higher impact for papers which were produced in collaboration it is not possible to use the expected value of 10% as a benchmark for the values in the excellence-networks.net. The average best paper rates given in each subject area should be used as benchmarks for the interpretation of the results in the application.

The user of the application should also consider the following point in the interpretation of the results: The best paper rates in excellence-networks.net are estimated on the basis of a statistical model. The reliability of the rate is taken into account in this model: when the number of papers written by the institutions in collaboration is smaller, the reliability of the observed best paper rate falls. The more unreliable the observed value for an institution, the more the best paper rate for an institution estimated from the observed values shrinks towards the overall mean value of a subject area. In other words, the overall mean value is the best estimate for the performance of an institution if the existing information about an institution is insufficiently reliable.

In this study, the assessments of the institutional collaboration activities are based on citation impact data only. This is only one indicator which measures performance in a specific way. There are many other performance indicators which can be used to assess these activities from other perspectives (e.g. jointly submitted grant applications or common patents). We plan to consider further indicators in next releases of the application.